\documentclass[aps,prl,showpacs,floatfix,twocolumn]{revtex4}
\usepackage{graphicx}
\usepackage{CJK}
\newcommand{\beq}{\begin{equation}}
\newcommand{\eeq}{\end{equation}}
\newcommand{\be}{\begin{eqnarray}}
\newcommand{\ee}{\end{eqnarray}}
\begin{document}
\begin{CJK*}{KS}{}
\title{Microcanonical analysis of a nonequilibrium phase transition}
\author{Julian Lee}
\email{jul@ssu.ac.kr}
\affiliation{Department of Bioinformatics and Life Science, Soongsil University, Seoul, Korea}
\date{\today}
\begin{abstract}
Microcanonical analysis is a powerful method for studying phase transitions of finite-size systems. This method has been used so far only for studying phase transitions of  equilibrium
systems, which can be described by microcanonical entropy. I show that it is possible to perform microcanonical analysis of a nonequilibrium phase transition, by generalizing the concept of microcanonical entropy. One-dimensional asymmetric diffusion process is studied as an example where such a generalized entropy can be explicitly found, and the microcanonical method is used to analyze a nonequilibrium phase transition of a finite-size system.
\end{abstract}
\maketitle
\end{CJK*}
Microcanonical analysis is a powerful method for studying phase transitions of finite-size systems~\cite{Gross1, Gross2}. In this approach, the form of the microcanonical entropy is examined to see whether there is a convex region. Existence of such a region signals the onset of an inhomogeneity, and the system is considered to undergo a first-order phase transition in this region. The phase transition in microcanonical analysis is a well-defined concept even for a finite-size system, which is in contrast to most of the traditional canonical ensemble approach where phase transition is defined only for infinite size systems, defined in terms of singular behavior of physical quantities in this limit. The microcanonical analysis has been applied for studying phase transitions of various finite-size systems such as spin models~\cite{spin1,spin2,spin3,spin4}, atomic clusters and nuclei~\cite{Gross2,frag}, polymers~\cite{pol1,pol2,pol3,pol4,pol5,pol6,pol7,pol8}, peptides~\cite{pep}, and proteins~\cite{pro1,pro2,pro3,pro4,pro5}.

 Since only equilibrium systems can be described in terms of microcanonical entropy, microcanonical method has been used exclusively for analyzing equilibrium phase transitions so far. In this Letter, I show that it is possible to apply this method for analyzing a nonequilibrium phase transition~\cite{open,closed,ASEP2,Ar00,pfz2,rev,park}, by a proper generalization of the concept of microcanonical entropy. 
 
 Let us first briefly review the connection between the convex region of microcanonical entropy and the phase transition~\cite{Gross1, Gross2}. We consider a finite closed system with a conserved quantity, say energy E, and denote the number of corresponding microstates as $\Omega_L(E)$, where the subscript denotes the dependence on the system size $L$. The microcanonical entropy is then defined as 
\be
S_L (E) = \ln \Omega_L(E)
\ee
where we use the unit with $k_B=1$. Now suppose we construct a larger system by assembling two identical subsystems of energy $E$ and size $L$. We let the two subsystems make a thermal contact, but let the coupling between the two systems be weak enough so that the total energy is $E_{\rm tot} = 2 E = E_A + E_B$ where $E_A$ and $E_B$ are the energy values of the two subsystems. We then examine the qualitative feature of the probability distribution of the energy values of the subsystems, $P(E_A, E_B) \propto\exp(S_L(E_A)+S_L (E_B))$. If  $S_L (E)$ is a concave function, then $P(E,E) >P(E_-,E_+)$ for any $E_\pm$ with $E_- < E < E_+$, so the homogeneous distribution of energy among the subsystems is preferred. On the other hand, if there is a convex region in $S_L(E)$ so that one can find values $E_1$, $E_2$ and $0< p < 1$ satisfying 
\be
E &=& p E_1 + (1-p) E_2,  \nonumber\\
S_L(E) &<& p S_L(E_1) + (1-p) S_L(E_2), \label{conv}
\ee
then there are values $E_\pm$ such that $P(E,E) < P(E_-,E_+)$ so that an inhomogeneous distributions is favored, and we say that the system is in the region of the first-order phase transition. The argument can easily be generalized to the case of subsystems of different sizes, more than two subsystems, and multiple conserved quantities~\cite{Gross1, Gross2}.  

We note  that the only relevant property of $\Omega_L(E)$ exploited in the argument is that when a conserved quantity ${\bf Q}={\bf Q_A}+{\bf Q_B}$ of the total system is distributed over two subsystems $A$ and $B$, the probability  distribution of ${\bf Q_A}$ and ${\bf Q_B}$ is proportional to the product of $\Omega$s:
\be
P ({\bf Q_A}, {\bf Q_B}) \propto \Omega_L({\bf Q_A})\Omega_{L'}({\bf Q_B}) \label{prod}
\ee 
where $L$ and $L'$ denote the sizes of the subsystems $A$ and $B$. We can also impose a certain boundary condition at the interface between the subsystems, in which case $P ({\bf Q_A}, {\bf Q_B})$ becomes a {\it conditional} probability. Therefore, it is clear that even for a nonequilibrium system,  if a  probability of the distribution of a conserved quantity  ${\bf Q}$ among subsystems under appropriate boundary condition can be expressed in the form Eq.(\ref{prod}), then we can consider $\Omega_L({\bf Q})$ as the generalized density, and their log as the generalized entropy, which can then be used as the target of the microcanonical analysis. This is the main claim of this Letter.

As an example of a nonequilibrium model on which microcanonical analysis can be performed, we consider a diffusion model where particles of two types, labelled as 1 and 2, move asymmetrically on a periodic lattice of length $L$~\cite{closed,ASEP2,Ar00,ASEP1}. Treating the vacancy as a particle with label $0$, the transition rates $g_{\alpha \beta}$ for the particle exchange of the type $(\alpha, \beta) \to (\beta, \alpha)$  at neighboring sites 
are given as~\cite{ASEP2,Ar00,ASEP1}
\be
g_{10}=g_{02}=1,\ g_{12}=q,\   g_{21}=1
\ee
with all other components of $g$s being zero. We note that the numbers of both types of particles are conserved separately, which we will denote as $n_1$ and $n_2$. The matrix representation of the stationary state for this process has already been found, and is given as~\cite{ASEP2,Ar00,ASEP1}
\be
P_{st}(\beta_1, \cdots \beta_L) = {\rm tr} {\bf G}_{\beta_1}\cdots {\bf G}_{\beta_L}/\sum_{\gamma_1,\cdots,\gamma_L} {\rm tr} {\bf G}_{\gamma_1}  \cdots {\bf G}_{\gamma_L} \label{stat}
\ee
where $\beta_k$ denotes the particle type at the $k$-the site, and the components of the three infinite-dimensional matrices ${\bf G}_\beta\ (\beta=0,1,2)$ are given as
\be
(G_0)_{ij} &=& \delta_{1i} \delta_{1j}\nonumber\\
(G_1)_{ij} &=& a_i \delta_{i j} + t_i \delta_{i j-1} \nonumber\\
(G_2)_{ij} &=& a_i \delta_{i j} + s_j \delta_{i-1 j}. \label{matrix}
\ee
where
\be
a_k &=& \frac{1+q^{2-k}-2q^{1-k}}{q-1}, \nonumber\\
s_k t_k  &=&  \frac{1-q^{-k}}{(q-1)^2} \left( 1 -q^{3-k} + 4 q^{2-k} - 4 q^{1-k}  \right)
\ee
Now let us suppose that there are vacancies at sites $a$ and $b$. The whole periodic lattice can be divided into two regions bounded by these two sites, and we would like to obtain conditional probability for particles in these two regions being ${\bf n}_A =(n_1^{(A)},n_2^{(A)})$ and ${\bf n}_B = (n_1^{(B)},n_2^{(B)})$.  Obviously, from Eq.(\ref{stat}), we see that it is proportional to
\be
&&P({\bf n}_A, {\bf n}_B) \propto \sum_{\gamma_1,\cdots,\gamma_L} {\rm tr} ( {\bf G}_{\gamma_1} \cdots {\bf G}_{\gamma_{a-1}} {\bf G}_0 {\bf G}_{\gamma_{a+1}} \cdots  {\bf G}_{\gamma_{b-1}}\nonumber\\
&& \times {\bf G}_0 {\bf G}_{\gamma_{b+1}} \cdots {\bf G}_{\gamma_L} ) \delta(\sum_{i\in A} \delta_{\gamma_i,1},n_1^{(A)}) \delta(\sum_{j\in A} \delta_{\gamma_j,2},n_2^{(A)} )\nonumber\\
&&\times \delta( \sum_{k\in B} \delta_{\gamma_k,1},n_1^{(B)} ) \delta( \sum_{l\in B} \delta_{\gamma_l,2},n_2^{(B)} )
\ee
where $\delta(a,b)=\delta_{a,b}$ denotes Kronecker delta function that vanishes when the indices are not equal.
Note that
\be
&&{\rm tr} ( {\bf G}_{\gamma_1} \cdots {\bf G}_{\gamma_{a-1}} {\bf G}_0  {\bf G}_{\gamma_{a+1}} \cdots {\bf G}_{\gamma_{b-1}} {\bf G}_0 {\bf G}_{\gamma_{b+1}} \cdots {\bf G}_{\gamma_L} )\nonumber\\
&=&{\rm tr} ( {\bf G}_0 {\bf G}_{\gamma_{a+1}} \cdots {\bf G}_{\gamma_{b-1}} {\bf G}_0 {\bf G}_{\gamma_{b+1}} \cdots   {\bf G}_{\gamma_{a-1}} )\nonumber\\
&=& \left[{\bf G}_{\gamma_{a+1}} \cdots  {\bf G}_{\gamma_{b-1}}\right]_{11} \left[{\bf G}_{\gamma_{b+1}} \cdots {\bf G}_{\gamma_{a-1}}\right]_{11}\nonumber\\
&=&{\rm tr} ( {\bf G}_0 {\bf G}_{\gamma_{a+1}} \cdots {\bf G}_{\gamma_{b-1}}){\rm tr} ({\bf G}_0 {\bf G}_{\gamma_{b+1}} \cdots   {\bf G}_{\gamma_{a-1}} ).
\ee
Therefore, the conditional probability for the steady state is expressed in the form Eq.(\ref{prod}), where the generalized density for a system of size $L$ is now defined as
\be
&&\Omega_L ({\bf n})\nonumber\\ &= & \sum_{\gamma_1,\cdots,\gamma_L} {\rm tr} ( {\bf G}_0 \prod_{k=1}^{L-1}{\bf G}_{\gamma_{k}}) \delta( \sum_{i=1}^{L-1} \delta_{\gamma_i,1},n_1 ) \delta( \sum_{j=1}^{L-1} \delta_{\gamma_j,2},n_2 )\nonumber\\
&=&  \sum_{\gamma_1,\cdots,\gamma_L} \left[\prod_{k=1}^{L-1} {\bf G}_{\gamma_{k}}\right]_{11}\!\!\!\!\! \delta( \sum_{i=1}^{L-1} \delta_{\gamma_i,1},n_1 ) \delta( \sum_{j=1}^{L-1} \delta_{\gamma_j,2},n_2), \label{gen}
\ee
where the system size $L$ includes one vacancy.
We analyze the phase transition of the current model by performing the microcanonical analysis on the generalized entropy $S_L({\bf n}) = \log \Omega_L ({\bf n})$. It is expressed in terms of $(L/2) \times (L/2)$ submatrices of ${\bf G}_\beta$, which can be computed exactly for given values of $q$ and $L$~\cite{ASEP2,Ar00,ASEP1}. 

By performing analytic computations, Monte Carlo simulation, mean field calculations~\cite{ASEP2}, and partition function zero analysis~\cite{Ar00}, it has been argued that this system undergoes a nonequilibrium phase transition in the limit of $L \to \infty$. There is a $q_c > 1$ such that the system remains homogeneous  for $q \ge q_c$, but inhomogeneities of particle densities appear for certain range of particle numbers when $q<q_c$. In fact, the latter can be considered as a region of first order transition between the fluid and the condensed phases, as will be elaborated below.
 
From the viewpoint of microcanonical analysis, the criterion for a first-order transition is the existence of a nonconcave region in the microcanonical entropy, a set of points where one can find a direction with positive second derivative~\cite{Gross1, Gross2}.  For the current model where the conserved quantity ${\bf n}$ is discrete, I examined discretized second derivatives
\be
\Delta_a \Delta_b S_L &\equiv& S_L(n_1+a, n_2+b)  +  S_L(n_1-a, n_2-b)\nonumber\\
&-& 2 S_L(n_1, n_2),
\ee
along the horizontal ($(a,b)=(1,0)$), vertical ($(a,b)=(0,1)$), and two diagonal ($(a,b)=(1,\pm 1)$) directions. A point in the interior is nonconcave if any one of these four quantities has a positive value. 

\begin{figure}
\includegraphics[width=\columnwidth]{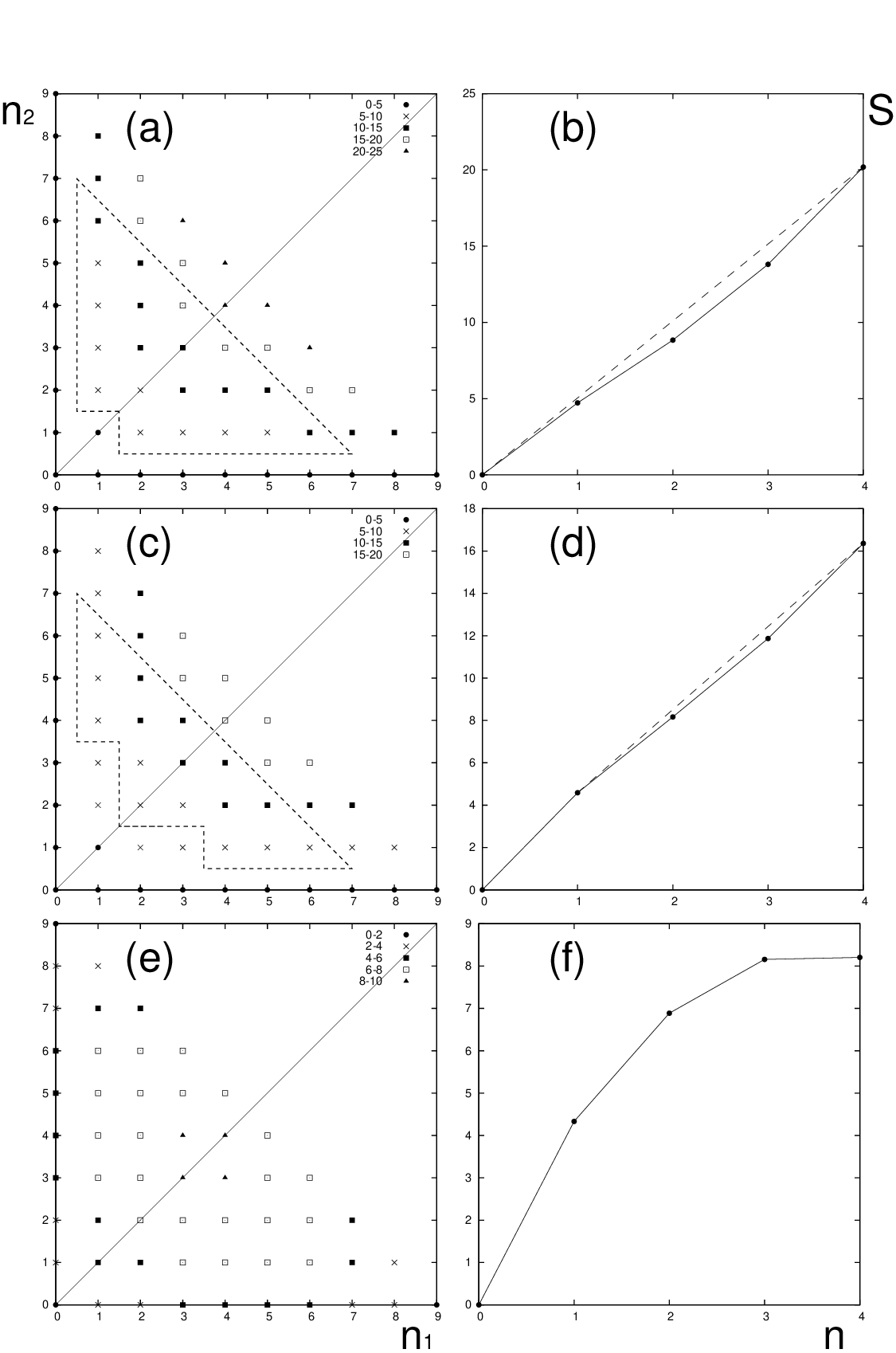}
\caption{The generalized entropy function $S_L(n_1, n_2)$ for $L=10$ is displayed with distinct symbols for points belonging to different ranges of function values, for (a) $q=0.5$ , (c) $q=0.7$, and (e) $q=2.0$.  The nonconcave region is enclosed by dashed lines. The cross section along the diagonal line $n_1=n_2$ in figures (a), (c), and (e), are displayed in figures (b), (d), and (f). The concave envelopes are drawn in figures (b) and (d) with dashed lines as visual guides.}
\label{d10} 
\end{figure}

\begin{figure}
\includegraphics[width=\columnwidth]{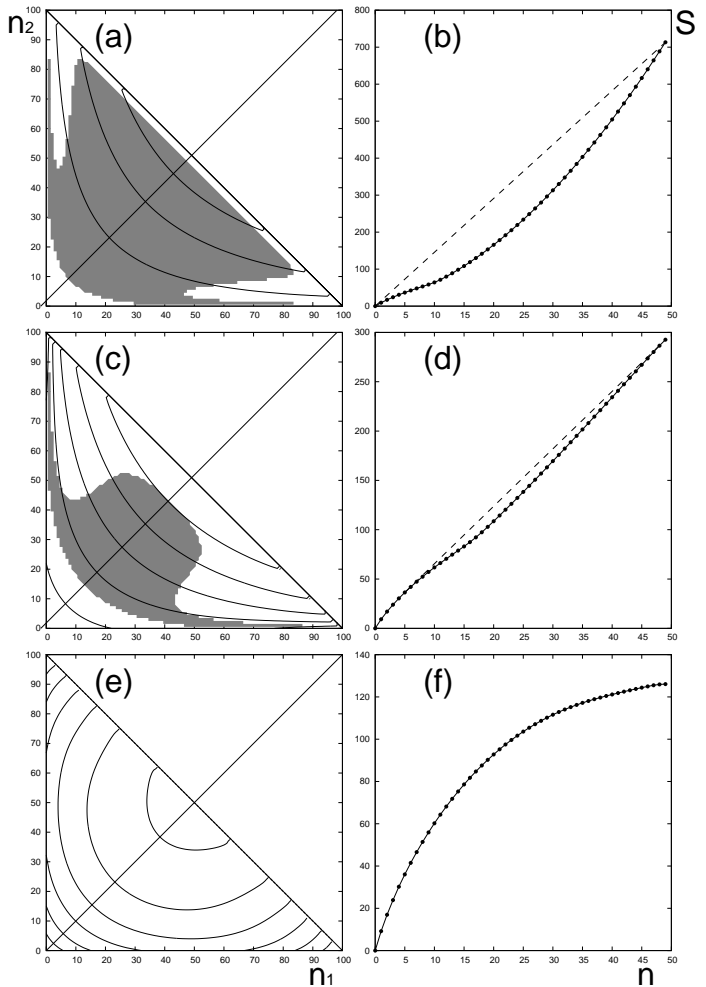}
\caption{The contours of the generalized entropy function $S_L(n_1, n_2)$ for $L=100$ are drawn for (a) $q=0.8$ , (c) $q=1.1$, and (e) $q=2.0$, at intervals of 200, 50, 20, respectively. The nonconcave region is colored gray. The cross section along the diagonal line $n_1=n_2$ in figures (a), (c), and (e), are displayed in figures (b), (d), and (f). The concave envelopes are drawn in figures (b) and (d) with dashed lines as visual guides.}
\label{d100} 
\end{figure}

The generalized entropy function  $S_L(n_1,n_2)$ are shown in the left panels of the figures \ref{d10} and \ref{d100} for $L=10$ and $L=100$ respectively, for various values of $q$. Note that the entropy has the symmetry with respect to the line $n_1=n_2$ due to the  invariance under the simultaneous application of particle type exchange $1 \leftrightarrow 2$ and the parity inversion $k \leftrightarrow -k$. We see that for small enough values of $q$, a nonconcave region appears in the generalized entropy, enclosed by dashed lines in figure \ref{d10} and denoted as gray regions in figure \ref{d100}.  As $q$ increases, the nonconcave region shrinks, and eventually disappears for large enough values of $q$. We find that nonconcave region always includes a part of the line $n_1=n_2$. The second derivative at such a point is also largest along the $(1,1)$ direction, which tells us that for sufficiently small $q$, when the system is divided into subsystems with respect to a pair of vacancies, it is most probable that there is a inhomogeneity for the total particle numbers, but there are the same numbers of two species at both sides. Note that this is an {\it exact} statement for a {\it finite} value of $L$, in contrast to the results of previous works where limit of $L \to \infty$ was considered~\cite{ASEP2, Ar00}. 

\begin{figure}
\includegraphics[width=\columnwidth]{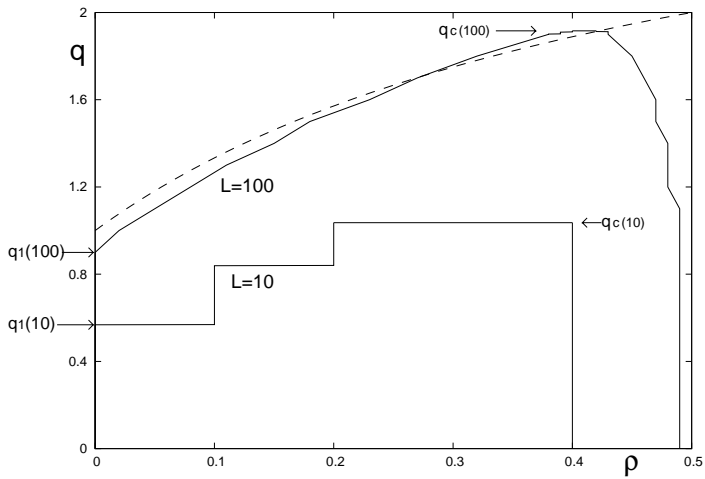}
\caption{The phase boundary $\rho_\pm (q)$ for $L=10$ and $L=100$. The mean field result in the limit of $L \to \infty$ is shown with dashed line for comparison.}
\label{Nq} 
\end{figure}

The cross sections of $S_L(n_1,n_2)$ along the line $n_1=n_2=n$, $S_L(n,n)$, are also displayed in the right panels of figures \ref{d10} and \ref{d100}, where the concave envelopes are denoted by dashed lines whenever they exist. These correspond to the regions of the first-order transition, whose upper and lower boundaries $\rho_\pm$ in the space of particle density $\rho \equiv n/L\ (0\ \le \rho < 0.5)$ are drawn as functions of $q$ to produce a phase diagram in the figure \ref{Nq}, for $L=10$ and $L=100$.  The mean field result $\tilde q(\rho) = (1 + 6 \rho)/(1 + 2 \rho)$ in the limit of $L \to \infty$ is shown in the figure \ref{Nq} with a dashed line for comparison~\cite{ASEP2},  where  $\tilde q(\rho)$ is the inverse function of  $\rho_\pm(q)$.  
 
The high-density side of the phase boundary, $\rho \ge \rho_+$, corresponds to the condensed phase. Note that there is a $q_1(L)$ such that $\rho_-=0$ for $q \le q_1(L)$, in which case the low-density phase $\rho=\rho_-=0$ is just the vacuum without any particles present. For $q>q_1(L)$, the low-density phase $\rho \le \rho_-$ is the fluid phase.  As $q$ increases, $\rho_\pm$ approaches toward each other and eventually merges at the critical point $q=q_c(L)$, after which the system is in a homogeneous phase. The values of $q_1$ and $q_c$ for $L=10$ and $L=100$ are indicated by arrows in Figure \ref{Nq}. The mean field prediction for these parameters are $q_1(\infty)=1$ and $q_c(\infty)=2$, as can be easily read off from the analytic expression for $\tilde q(\rho)$.

The regions $q \le q_1$, $q_1 < q < q_c$, and $q \ge q_c$ have been called pure, mixed, and disordered phases~\cite{ASEP2}. However,  microcanonical analysis shows that in $\rho$ space, each of the regions $q \le q_c$ and $q_1 < q < q_c$  is divided into vacuum (or fluid) phase ($\rho \le \rho_-$), condensed phase ($\rho \ge \rho_+$),  and the phase coexistence region ($\rho_- < \rho < \rho_+$). 
The situation is analogous to the  two-dimensional Ising model with the conserved magnetization $M$ and the temperature $T$. When one simply considers the $T$ dependence, then there is a critical temperature $T_c$ such that the system is in a disordered phase for $T \ge T_c$ and ordered phase for $T < T_c$. However, by examining the $M$ dependent behavior of the system, one realizes that the ordered phase in fact gets divided into up-spin phase, down-spin phase, and the region of the first-order transition between up and down phases. 

\begin{figure}
\includegraphics[width=\columnwidth]{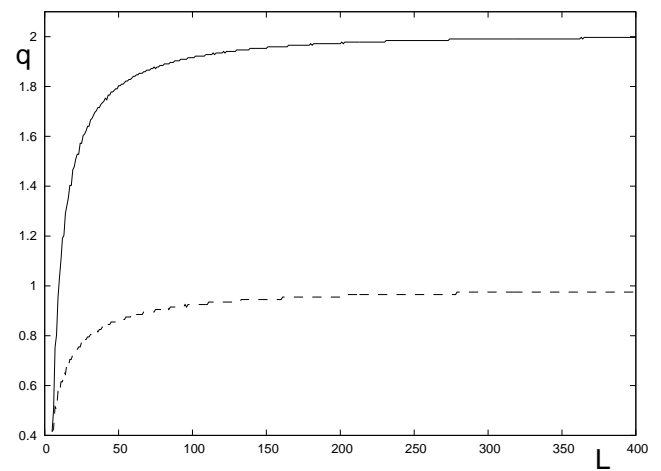}
\caption{The critical values $q_c$ (solid line) and $q_1$ (dashed line) as functions of the system size $L$. The first-order phase transition exists for $q<q_c$. The low-density phase is a vacuum phase for $q \le q_1$ and fluid phase for $q_1<q<q_c$ }
\label{phase} 
\end{figure}

I also plot $q_c(L)$ and $q_1(L)$ as functions of $L$ in figure \ref{phase}. Both $q_c(L)$ and $q_1(L)$ approach their mean field values $q_c(\infty)=2$ and $q_1(\infty)=1$.

 The current work also clarifies the physical meaning of a previous work based on the partition function zeros (PFZs)~\cite{Ar00}. There, a partition function of the form $\sum_{n_1,n_2}\Omega_L(n_1,n_2) x^{n_1+n_2}$ was constructed where $x$ was called the chemical potential. Then the PFZs in the complex plane of $x$ was analyzed to claim that there is a first-order transition as $L \to \infty$, for sufficiently small values of $q$. It is obvious that $\Omega_L({\bf n})$ was used implicitly as the generalized density of states, but it was not explained why $\Omega_L({\bf n})$ should have such a special status. Also, the physical meaning of the chemical potential was unclear, because $\Omega_L({\bf n})$ was regarded as describing the  the particles on a periodic lattice of size $L$, which is an isolated system. The current work not only justifies the use of $\Omega_L({\bf n})$ as a generalized density, via the factorization Eq.(\ref{prod}), but also shows that $\Omega_L({\bf n})$ in PFZs approach  describes a subsystem of size $L-1$ bounded by pair of vacancies, rather than the whole system. Then the chemical potential $x$ can be considered as a parameter describing the rest of the system whose size is much larger than $L$, acting as an infinite-size particle reservoir.  The microcanonical analysis is more general since the phase transition is well defined for a system with a finite size. In fact, the notion of finite-size nonequilibrium phase transition itself is introduced for the first time in the current work via microcanonical analysis, which would be a subject of much interest for future study. 

This work was supported by the National Research Foundation of Korea, funded by the Ministry of Education, Science, and Technology (NRF-2014R1A1A2058188).

\end{document}